\documentclass[pre,floatfix,notitlepage,nofootinbib]{revtex4-2}

\usepackage[utf8]{inputenc}
\usepackage[T2A]{fontenc}

\usepackage{amsmath}
\usepackage{amssymb}
\usepackage{lipsum}
\usepackage{bm}
\usepackage{graphicx}
\usepackage{graphics}
\usepackage[dvipdf]{epsfig}
\usepackage{subcaption}
\usepackage{float}
\usepackage{wrapfig}
\usepackage{url}
\usepackage{epstopdf} 
\usepackage{xcolor}
\usepackage{tensor} 
\usepackage{setspace}
\usepackage{commath} 
\usepackage{newtxmath}

\newcommand{\bx}{{\mathbf x}}
\newcommand{\bX}{{\mathbf X}}
\newcommand{\bg}{{\mathbf g}}
\newcommand{\bG}{{\mathbf G}}
\newcommand{\bI}{{\mathbf I}}
\newcommand{\ba}{{\mathbf a}}
\newcommand{\bA}{{\mathbf A}}
\newcommand{\bn}{{\mathbf n}}
\newcommand{\bN}{{\mathbf N}}
\newcommand{\bb}{{\mathbf b}}

\newcommand{\bEB}{\mathbf{E}_{\text{B}}}
\newcommand{\bSB}{\boldsymbol\Sigma_{\text{Biot}}}
\newcommand{\bU}{{\mathbf{U}}}
\newcommand{\bV}{{\mathbf{V}}}
\newcommand{\bC}{{\mathbf{C}}}
\newcommand{\bB}{{\mathbf{B}}}
\newcommand{\bQ}{{\mathbf{Q}}}
\newcommand{\ant}{\text{\Large{\bf{\cyrl}}}}

\begin{document}

\title{Energies for elastic plates and shells from quadratic-stretch elasticity}
\author{E. Vitral}
	\email{evitral@unr.edu}
\author{J. A. Hanna}
	\email{jhanna@unr.edu}
\affiliation{Department of Mechanical Engineering, University of Nevada,
     1664 N.\ Virginia St.\ (0312), Reno, NV  89557-0312, U.S.A.}

 \begin{abstract}
 We derive stretching and bending energies for isotropic elastic plates and shells. 
Through the dimensional reduction of a bulk elastic energy quadratic in Biot strains, we obtain two-dimensional bending energies quadratic in bending measures featuring a bilinear coupling of stretches and geometric curvatures.
 For plates, the bending measure is invariant under spatial dilations and naturally extends primitive bending strains for straight rods. 
For shells or naturally-curved rods, the measure is not dilation invariant, and contrasts with previous \emph{ad hoc} postulated forms.
The corresponding field equations and boundary conditions feature moments linear in the bending measures, and a decoupling of stretching and bending such that application of a pure moment results in isometric deformation of a unique neutral surface, primitive behaviors in agreement with classical linear response but not displayed by commonly used analytical models. 
We briefly comment on relations between our energies, those derived from a neo-Hookean bulk energy, and a commonly used discrete model for flat membranes. 
Although the derivation requires consideration of stretch and rotation fields, the resulting energy and field equations can be expressed entirely in terms of metric and curvature components of deformed and reference surfaces. 
\end{abstract}

\date{\today}


\maketitle



\section{Introduction}

This paper provides a detailed derivation of bending measures and energies for elastic plates and shells that were proposed in a companion paper \cite{vitral2021dilation} through physical arguments related to a sensible definition of ``pure stretching'' of a surface, and whose properties and advantages were discussed therein. 
The present results follow from dimensional reduction of a general isotropic bulk elastic energy quadratic in Biot strains \cite{john1960plane,lurie1968theory,vitral2021quadratic}. 
The use of such strains as primitive quantities in elasticity emerged in the field of rod mechanics \cite{IrschikGerstmayr09}, and 
its advantages for small-strain elasticity theories in soft matter were recently discussed in great detail by Oshri and Diamant \cite{OshriDiamant17} and subsequently by Wood and Hanna \cite{WoodHanna19}, both in the context of thin bodies. 
These advantages may be summarized as follows.  A bending energy quadratic in an appropriate primitive measure will give rise to a moment linear in the measure and a tangential force without extra bending contributions, which aside from its appealing simplicity has the important consequence that pure moments will not induce stretching or compression of a neutral surface; for a plate, this is the mid-surface. 
 A choice of energy is a choice of constitutive relations, with the potential to couple the stretching and bending response of a structure in a variety of simple settings; induction of stretching by moments may occlude fundamental questions regarding the geometric coupling of stretching and bending, particularly as may occur in elastic singularities. 
 The present paper and its companion \cite{vitral2021dilation} offer an extension of primitive bending measures for straight rods and axisymmetric plates to classes of thin elastic bodies including general isotropic plates, curved rods, and shells.  
These measures may serve as ideal building blocks for theories of thin structures.
The plate measure can be seen as a symmetrized form of a tensor suggested by Atluri \cite{atluri1984alternate}, while those for curved rods and shells that we derive here, and also physically justify in \cite{vitral2021dilation}, differ from the many \emph{ad hoc} postulated forms in the literature. 
  
  The basis of the present derivation is a bulk elasticity theory formed by systematic expansion in Biot strains or similar quantities linear in stretch \cite{vitral2021quadratic}, including low-order terms not present in neo-Hookean elasticity or theories constructed from metric differences (Green-Lagrange or Euler-Almansi strains). 
The use of stretch tensors requires consideration of a rotation tensor or tensor square root operations, and in three dimensions these introduce significant complications not present when using metric components.  
However, simplifications are possible in the setting of a two-dimensional surface, with all the relevant quantities eventually related to metric and curvature components such that the rotations need never be explicitly found.

The dimensional reduction we follow employs a generalized Kirchhoff-Love kinematics \cite{StumpfMakowski86, ozenda2021kirchhoff} that 
avoids the internal inconsistencies of the classical assumptions, while retaining the flexibility to not presuppose any particular scaling of stretching or bending with thickness, as has been done in more rigorous mathematical approaches. 
We combine this with an approach to constructing effective shell field theories from a series of papers by Steigmann~\cite{steigmann2007thin,steigmann2008two,steigmann2012extension,steigmann2013koiter}, while taking care to compute through-thickness derivatives of stretches in a manifestly symmetric way. 
  
The structure of the paper is as follows.  Notation, the basic kinematics of plates and shells, and important tensors such as stretches and curvatures, are introduced in Section \ref{sec:def}.
Quadratic-stretch elasticity and its associated energy are reviewed in Section \ref{sec:biot}.
Section \ref{sec:reduction} applies a dimensional reduction procedure to this energy. Section \ref{sec:energy} presents the resulting two-dimensional stretching and bending energies for shells, the latter based on a new tensor measure of bending.  Plates, whose bending energy is dilation-invariant, and curved rods are discussed as special cases. Comparison is made with a commonly used discrete bending energy.  Field equations and boundary conditions are presented in Section \ref{sec:field}.  The linearity of the force and moment and the consequences for deformation under simple loadings are demonstrated, including recovery of the unique neutral surface of classical linear elasticity. 
Finally, Section \ref{sec:determination} replaces the description in terms of stretches and rotations with another in terms of deformed and referential metric and curvature components.


\section{Notation, kinematics, stretch, rotation, and curvature}
\label{sec:def}

We adopt the following notation.
 $\textrm{Tr}\,\mathbf{D}$ and $\textrm{Det}\,\mathbf{D}$ are the trace and determinant of a tensor $\mathbf{D}$, whereas $\textrm{det} [D_{ij}]$ is the determinant of a matrix with entries $D_{ij}$.
$\textrm{sym}\,\mathbf{D} = \frac{1}{2}(\mathbf{D}+\mathbf{D}^\top)$ is the symmetric part of a second order tensor $\mathbf{D}$ whose transpose is $\mathbf{D}^\top$.  $\| \mathbf{D} \|$ is an (unspecified) tensor norm.  We will employ the shorthands $\mathrm{Tr}^2() \equiv \left[ \mathrm{Tr}\,()\right]^2$, $\mathrm{Tr}\,()^2 \equiv \mathrm{Tr}\left[ ()^2 \right]$, and $\mathrm{sym}^2() \equiv \left[ \mathrm{sym}\,()\right]^2$, $\mathrm{sym}\,()^2 \equiv \mathrm{sym}\left[ ()^2 \right]$.
 
An elastic body $\mathcal{B}$ of reference thickness $h$ is parameterized by a system of material coordinates $\{\eta^\alpha,\zeta\}$, 
where $-h/2 \leq \zeta \leq h/2$ is a coordinate indicating distance from a mid-surface $\mathcal{S}$; in the kinematics we consider, it will be a normal coordinate and $h$ will be presumed small with respect to other relevant length scales. 
 Latin indices run over all three material coordinates, while Greek indices run over the two lateral coordinates $\eta^\alpha$.
We consider deformations between a referential (rest) embedding $\mathbf{R}(\eta^\alpha,\zeta)$ with mid-surface $\mathbf{X}(\eta^\alpha) = \mathbf{R}(\eta^\alpha,0)$ and a present (deformed) embedding $\mathbf{r}(\eta^\alpha,\zeta)$ with mid-surface $\mathbf{x}(\eta^\alpha) = \mathbf{r}(\eta^\alpha,0)$, all in $\mathbb{E}^3$. 
We denote (non-covariant) material derivatives with a subscripted $d$, however, following common practice and ensuring compactness of derivations, we will often denote the through-thickness $\zeta$-derivatives with a prime $'$.
The embeddings $\mathbf{R}$ and $\mathbf{r}$ have, respectively, coordinate bases in the form of tangents $\mathbf{G}_i = d_i\mathbf{R}$ and $\mathbf{g}_i = d_i\mathbf{r}$ and reciprocal tangents defined through the relations $\mathbf{G}^i\cdot\mathbf{G}_j = \mathbf{g}^i\cdot\mathbf{g}_j = \delta^i_j$. 
The mid-surfaces $\bX$ and $\bx$ have, respectively, unit normals $\bN$ and $\bn$, tangents $\bA_\alpha = d_\alpha\bX$ and $\ba_\alpha = d_\alpha\bx$, and reciprocal tangents defined through the relations $\bA^\alpha\cdot\bA_\beta = \ba^\alpha\cdot\ba_\beta = \delta^\alpha_\beta$.
Metric and inverse metric components on the referential and present bodies and mid-surfaces follow naturally: 
$G_{ij} = \mathbf{G}_i\cdot\mathbf{G}_j$, $G^{ij} = \bG^i\cdot\bG^j$, 
$g_{ij} = \mathbf{g}_i\cdot\mathbf{g}_j$, $g^{ij} = \bg^i\cdot\bg^j$, 
$A_{\alpha\beta} = \mathbf{A}_\alpha\cdot\mathbf{A}_\beta$, $A^{\alpha\beta} = \bA^\alpha\cdot\bA^\beta$, 
$a_{\alpha\beta} = \mathbf{a}_\alpha\cdot\mathbf{a}_\beta$, $a^{\alpha\beta}= \ba^\alpha\cdot\ba^\beta$.  
These can be applied to raise or lower indices only on those objects corresponding to the same manifold as the metric in question.
We define referential and present surface metric determinants $A = \textrm{det}[A_{ij}]$
and $a = \textrm{det}[a_{ij}]$, surface covariant derivatives $\bar\nabla_\alpha$ and $\nabla_\alpha$, and 
surface gradients 
$\bar{\nabla} \equiv d_\alpha(\,)\mathbf{A}^\alpha$ and $\nabla \equiv d_\alpha(\,)\mathbf{a}^\alpha$. 
The symmetric curvature tensors of the referential and present surfaces are 
$\bar{\mathbf{b}} = \bar b_{\alpha\beta}\bA^\alpha\bA^\beta= -\bar{\nabla}\mathbf{N}$
and $\mathbf{b} = b_{\alpha\beta}\ba^\alpha\ba^\beta= -\nabla\mathbf{n}$, where $\bar b_{\alpha\beta} = d_\beta \bA_\alpha\cdot\bN = -\bA_\alpha\cdot d_\beta\bN$ and $b_{\alpha\beta} = d_\beta \ba_\alpha\cdot\bn = -\ba_\alpha\cdot d_\beta\bn$. 
The surfaces' mean and Gau{\ss}ian curvatures are invariants, $\bar{H} = \tfrac{1}{2}\textrm{Tr}\,\bar{\mathbf{b}} = \tfrac{1}{2}\bar b^\alpha_\alpha$, $\bar{K} = \textrm{Det}\,\bar{\mathbf{b}} = \tfrac{1}{2}\left(\bar b^\alpha_\alpha \bar b^\beta_\beta - \bar b^\alpha_\beta\bar b^\beta_\alpha\right)$,  $H = \tfrac{1}{2}\textrm{Tr}\,\mathbf{b} = \tfrac{1}{2} b^\alpha_\alpha$, $K = \textrm{Det}\,\mathbf{b} = \tfrac{1}{2}\left( b^\alpha_\alpha  b^\beta_\beta -  b^\alpha_\beta b^\beta_\alpha\right)$,
where of course $\bar b^\alpha_\beta = A^{\alpha\gamma}\bar b_{\gamma\beta}$ and $ b^\alpha_\beta = a^{\alpha\gamma} b_{\gamma\beta}$.

The reference and deformed configurations are taken to be of the form
\begin{align}
    \mathbf{R}(\eta^\alpha,\zeta) &= \mathbf{X}(\eta^\alpha)+\zeta\mathbf{N}(\eta^\alpha) \,, \label{eq:plc} \\
    \mathbf{r}(\eta^\alpha,\zeta) &= \mathbf{x}(\eta^\alpha)+\phi(\eta^\alpha,\zeta)\,\mathbf{n}(\eta^\alpha)\,,
    \label{eq:def}
\end{align}
with $\mathbf{N}$ constant for plates.
Defining these forms is equivalent to prescribing a deformation mapping that maps the reference position $\mathbf{R}$ of any material particle to a deformed position $\mathbf{r}$. 
The classical Kirchhoff-Love hypotheses assume $\phi = \zeta$, and thus do not allow changes in thickness whose associated stretching energetics need to be accounted for through another step in the calculation that is inconsistent with the kinematics. 
The generalized kinematics \eqref{eq:def} maintains the assumption that material fibers normal to the reference surface remain so under deformation, but allows for changes in thickness.
One of the first appearances of this generalized theory in the nonlinear deformations of shells may be found in Biricikoglu and Kalnins~\cite{biricikoglu1971large}, who considered the function $\phi$ to be linear in $\zeta$. Later, Chernykh~\cite{chernykh1980nonlinear} as well as Stumpf and Makowski~\cite{StumpfMakowski86} explored quadratic and more general dependence of $\phi$ on $\zeta$, arguing that the distribution of normal strains in the normal direction is essential for modeling large rotations and deformations of shells. 
While the complexity of the resulting field equations increases with the polynomial order of $\phi$, its coefficients can be easily determined when constraints on strains (or stresses), such as incompressibility, are imposed. Recently, Ozenda and Virga~\cite{ozenda2021kirchhoff} revisited the generalized Kirchhoff-Love theory, proposing that the leading coefficients of an expanded $\phi$ in $\zeta$ for compressible materials can be obtained by minimization of stretching and bending energies. 
Following this work, we expand 
\begin{equation}
    \phi(\eta^\alpha,\zeta) = \alpha_1(\eta^\alpha)\zeta+\alpha_2(\eta^\alpha)\zeta^2 + O(\zeta^3) \,,
    \label{eq:phi}
\end{equation}
with classical Kirchhoff-Love having $\alpha_1 = 1$ and all other coefficients zero. 
For compressible materials, the coefficients $\alpha_k(\eta^\alpha)$ can be determined either by energy minimization~\cite{ozenda2021kirchhoff} or by applying three-dimensional field equations and boundary conditions on the top and bottom of the shell \cite{steigmann2013koiter}; in either case, the form of $\phi$ will be material-dependent. 
For incompressible materials, the coefficients are determined directly by the kinematic constraint, and $\phi$ is material-agnostic. 
We further assume that $\phi'(\eta^\alpha,0) = \alpha_1(\eta^\alpha) > 0$; recall that a prime $'$ denotes differentiation with respect to $\zeta$. 
As shown in~\cite{ozenda2021kirchhoff} in the context of plates, 
this assumption means that for sufficiently small thickness $h$, 
the through-thickness derivatives dominate the surface derivatives of $\phi$, so that 
$|d_\beta\phi(\eta^\alpha,\zeta)\bG^\beta| \ll \phi'(\eta^\alpha,\zeta)$ everywhere.  This allows a simplifying approximation of the deformation gradient, 
\begin{equation}
    d_i\mathbf{r}\bG^i = \bg_i\bG^i \approx \bar{\nabla}\mathbf{x}+\phi\bar{\nabla}\mathbf{n}+\phi'\mathbf{n}\mathbf{N} \,,
    \label{eq:dg}
\end{equation}
where $\bar{\nabla}\mathbf{x} = d_\alpha\bx\bA^\alpha = \mathbf{a}_\alpha\mathbf{A}^\alpha$ can be thought of as a two-dimensional deformation gradient for the mid-surface. These ``two-point'' tensors are elements of the tensor product between the tangent spaces of the referential and present manifolds. 

The polar decomposition of the deformation gradient is 
\begin{align}
	\bg_i\bG^i = \bQ\cdot\vmathbb{U} = \vmathbb{V}\cdot\bQ\,,\label{eq:polar3d}
\end{align}
where the two-point rotation tensor $\bQ \in$ SO(3) and the symmetric positive-definite right (referential) $\vmathbb{U}$ and left (present) $\vmathbb{V}$ stretch tensors are three-dimensional.   These tensors encode the ratio of deformed length to rest length, reducing to the identity $\bI$ in an unstrained body. 
Through a justifiable abuse of notation, it is also possible \cite{pietraszkiewicz2008determination} to decompose the two-dimensional deformation gradient using  
\begin{equation}
  \ba_\alpha\bA^\alpha = \bQ\cdot\bU = \bV\cdot\bQ\,,
  \label{eq:polar}
\end{equation}
where $\bU$ and $\bV$ are two-dimensional, respectively restricted to the reference and present mid-surfaces where $\zeta=0$, and the three-dimensional rotation $\bQ$ is evaluated on the mid-surface.
The rotation transforms unit normals according to $\bQ\cdot\bN = \bN\cdot\bQ^\top = \bn$ and $\bQ^\top\cdot\bn=\bn\cdot\bQ =\bN$.  Thus, $\bQ'\cdot\mathbf{N} = \bN\cdot\left(\bQ^\top\right)' = \left(\bQ^\top\right)'\cdot\bn = \mathbf{n}\cdot\bQ' = \mathbf{0}$.
 In keeping with a generalized Kirchhoff-Love kinematics in which normals remain normal, we will further assume that the rotation does not vary significantly through the thickness, so that
$\|\bQ'\| |_{\zeta=0}$ is at most quadratic in a suitably normalized thickness and, thus, 
the $\zeta$-derivative of the deformation gradient, used later to derive the bending energy, has the form $(\bg_i\bG^i)'|_{\zeta=0} \approx \bQ\cdot\vmathbb{U}'|_{\zeta=0}$~\cite{wisniewski1998shell}. 
This is in line with the assumptions on $\|\bQ'\|$ adopted in~\cite{sansour1992exact,wisniewski2010finite}.

While, throughout this paper, we will be making use of the stretch tensors rather than the more familiar right $\vmathbb{C}$ and left $\vmathbb{B}$ Cauchy-Green deformation tensors, calculation of the latter lends insight into the form of the former.
Using the approximation~\eqref{eq:dg} of the deformation gradient and partially adopting the notation of \cite{ozenda2021kirchhoff}, we find
\begin{align}
  \vmathbb{C} &= (\bg_i\bG^i)^\top\cdot\bg_i\bG^i =  g_{ij}\mathbf{G}^i\mathbf{G}^j = \bC_\phi + (\phi')^2\mathbf{N}\mathbf{N}\,,
  \label{eq:c} \\
  \bC_\phi &= \bC + \phi(\bar{\nabla}\mathbf{x}^\top\cdot\bar{\nabla}\mathbf{n}+\bar{\nabla}\mathbf{n}^\top\cdot\bar{\nabla}\mathbf{x}) +\phi^2\bar{\nabla}\mathbf{n}^\top\cdot\bar{\nabla}\mathbf{n} \,,
  \label{eq:cphi} \\
  \bC &= \bar{\nabla}\mathbf{x}^\top\cdot\bar{\nabla}\mathbf{x} = a_{\alpha\beta}\mathbf{A}^\alpha\mathbf{A}^\beta \,,
  \label{eq:c2d} \\
  \vmathbb{B} &= \bg_i\bG^i \cdot (\bg_i\bG^i)^\top = G^{ij}\bg_i\bg_j = \bB_\phi + (\phi')^2\mathbf{n}\mathbf{n}\,,
  \label{eq:b} \\
  \bB_\phi &= \bB + \phi(\bar{\nabla}\mathbf{x}\cdot\bar{\nabla}\mathbf{n}^\top+\bar{\nabla}\mathbf{n}\cdot\bar{\nabla}\mathbf{x}^\top) +\phi^2\bar{\nabla}\mathbf{n}\cdot\bar{\nabla}\mathbf{n}^\top \,,
  \label{eq:bphi} \\
  \bB &= \bar{\nabla}\mathbf{x}\cdot\bar{\nabla}\mathbf{x}^\top = A^{\alpha\beta}\ba_\alpha\ba_\beta \,,
  \label{eq:b2d} 
\end{align}
where $\bC_\phi$, $\bC$, $\bB_\phi$, $\bB$ are two-dimensional tensors. The subscript $\phi$ on $\bC_\phi$ and $\bB_\phi$ indicates their dependence on the normal coordinate $\zeta$ through the function $\phi$, whereas $\bC$ and $\bB$ are respectively equal to $\bC_\phi$ and $\bB_\phi$ at $\zeta=0$. 
We may rewrite \eqref{eq:cphi} and \eqref{eq:bphi} in terms of the curvature tensor of the deformed surface
\begin{align}
  \bC_\phi &= \bC - 2\phi\bar{\nabla}\mathbf{x}^\top\cdot\mathbf{b}\cdot\bar{\nabla}\mathbf{x} +\phi^2\bar{\nabla}\mathbf{x}^\top\cdot\mathbf{b}^2\cdot\bar{\nabla}\mathbf{x} \,, \label{eq:cphib} \\
  \bB_\phi &= \bB - \phi\left(\bB\cdot\bb + \bb\cdot\bB\right) +\phi^2\bb\cdot\bB\cdot\bb \,. \label{eq:bphib}
\end{align}

 As $\bQ^\top\cdot\bQ = \bI = \bQ\cdot\bQ^\top$ is the three-dimensional identity, from \eqref{eq:polar3d}  we see that $\vmathbb{C} = \vmathbb{U}^2$ and $\vmathbb{B} = \vmathbb{V}^2$. 
 Then~\eqref{eq:c} and~\eqref{eq:b} imply that the right and left stretches are of the similar form
\begin{align}
    \vmathbb{U} &= \bU_\phi+\phi'\mathbf{N}\mathbf{N}\,,
    \label{eq:u} \\
    \vmathbb{V} &= \bV_\phi+\phi'\mathbf{n}\mathbf{n}\,,
    \label{eq:v}
\end{align}
with $\bU$ and $\bV$ from \eqref{eq:polar} respectively equal to $\bU_\phi$ and $\bV_\phi$ at $\zeta = 0$.


\section{Quadratic-Biot elastic theory}
\label{sec:biot}

The Cauchy-Green deformation tensors and their associated (Green-Lagrange or Euler-Almansi) strains involving metric differences are already quadratic in stretch, meaning that an energy quadratic in these strains will be quartic in stretch.  
In order to construct our primitive bending energy, we will instead employ a strain measure linear in stretch, guided by previous works \cite{IrschikGerstmayr09, OshriDiamant17, WoodHanna19, vitral2021quadratic}.
This measure is the Biot strain $\bEB = \vmathbb{U}-\mathbf{I}$, a referential tensor whose present counterpart, with the same eigenvalues, is the Bell strain $\vmathbb{V}-\mathbf{I}$.  Note that these three-dimensional quantities, which we will evaluate on the surface, should not be confused with the corresponding surface Biot and Bell strains $\bU - \bA_\alpha\bA^\alpha$ and $\bV - \ba_\alpha\ba^\alpha$ that will appear later in our reduced energy. 
  Our use of Biot rather than Bell is an arbitrary choice that makes use of prior derivations \cite{vitral2021quadratic}, however, when we wish to use the curvature tensor of the deformed surface in our description of bending, we will find that the left surface stretch $\mathbf{V}$ appears naturally. 

The principal invariants of $\bEB$ are
\begin{align}
  i_1^{\bEB} &= \quad\; \textrm{Tr}\,\bEB \qquad\qquad\;\;\; = \Delta_1+\Delta_2+\Delta_3\,,\nonumber
  \\
  i_2^{\bEB} &= \frac{1}{2}\left(\textrm{Tr}^2\bEB-\textrm{Tr}\,\bEB^2\right) = \Delta_1\Delta_2+\Delta_2\Delta_3+\Delta_1\Delta_3\,,
  \\
  i_3^{\bEB} &= \quad\; \textrm{Det}\,\bEB \qquad\qquad\; = \Delta_1\Delta_2\Delta_3\,,\nonumber
\end{align}
where the eigenvalues $\Delta_k$ are the principal Biot strains. 
The most general isotropic quadratic energy density in terms of the symmetric strain $\bEB$ is~\cite{vitral2021quadratic}
\begin{equation}
    \mathcal{W}(\bEB) = c_1(i_1^{\bEB})^2+c_2i_2^{\bEB}\,, 
    \label{eq:biot}
\end{equation}
also known as a ``semilinear'' material \cite{lurie1968theory} or, in two dimensions, a ``harmonic'' material \cite{john1960plane}.  
For $\mathcal{W}$ to be positive definite, the constant material parameters satisfy $c_1 \geq -c_2/3$ and $c_2 \leq 0$.\footnote{Note that the bounds on these coefficients mentioned in \cite{vitral2021quadratic} were overly conservative.}  
The conjugate quantity to the Biot strain is the symmetric Biot stress 
\begin{equation}
    \bSB = \frac{\partial \mathcal{W}}{\partial \bEB} = (2c_1+c_2)i_1^{\bEB}\mathbf{I}-c_2\bEB \,.
    \label{eq:sigma}
\end{equation}

From \eqref{eq:u} we may see that on the mid-surface ($\zeta = 0$), the Biot strain is $\bU+\alpha_1\mathbf{N}\mathbf{N}-\mathbf{I}$. Expanding the three-dimensional Biot strain in $\zeta$, 
\begin{equation}
    \bEB= 
    \left(\bU+\alpha_1\mathbf{N}\mathbf{N}-\mathbf{I}\right) +\zeta\vmathbb{U}'\big|_{\zeta=0}+\tfrac{1}{2}\zeta^2\vmathbb{U}''\big|_{\zeta=0}+O(\tilde h^3)\,,
    \label{eq:biot-exp}
\end{equation}
where we interpret $\tilde h$ as being the thickness suitably normalized by a length scale related to the derivative of the stretch, a quantity that does not in general coincide with the geometric curvature. 
The first term in \eqref{eq:biot-exp} is of the order of a characteristic Biot strain $\Delta$.  The resulting quadratic energy density we obtain will neglect terms of $O(\Delta^3, \tilde h\Delta^2, \tilde h^2\Delta, \tilde h^3)$.  Note that integration over the thickness will result in a factor of (non-normalized) $h$ multiplying everything, and the volume form for a shell will multiply by additional terms containing $h$ normalized by rest curvatures, which are related to the relative shallowness of the shell rather than the bending deformations affecting $\tilde h$.


\section{Reduction}
\label{sec:reduction}

Our descent into two dimensions will be ferried by Steigmann~\cite{steigmann2007thin,steigmann2008two,steigmann2012extension,steigmann2013koiter}, whose procedure we employ throughout the present section. 
However, we provide certain twists leading to new results.
In particular, the treatment of the through-thickness derivative of the stretch is delicate and requires care to avoid introducing artificial asymmetries into the resulting bending measures.\footnote{We don't want to end up with ``odd bending elasticity''.} 

The total elastic energy of the body $\mathcal{B}$ is
\begin{equation}
    \mathcal{E} = \int_\mathcal{B} \!\textrm{d}V\,\mathcal{W}\left[\bEB(\eta^\alpha,\zeta)\right] = \int_\mathcal{S}\int^{h/2}_{-h/2}\!\textrm{d}\zeta\,\textrm{d}\eta^1\textrm{d}\eta^2\sqrt{A}\,(1 - 2\zeta\bar{H}+\zeta^2\bar{K})\,\mathcal{W}\left[\bEB(\eta^\alpha,\zeta)\right] \equiv \int_\mathcal{S}\!\textrm{d}A\, w(\eta^\alpha) \,,
\end{equation}
where
$\textrm{d}V = \textrm{d}\zeta\,\textrm{d}A\,(1 - 2\zeta\bar{H}+\zeta^2\bar{K})$ is the referential volume form of the body and $\textrm{d}A = \sqrt{A}\,\textrm{d}\eta^1\textrm{d}\eta^2$ is the referential area form of its mid-surface. 
For three-dimensional energy densities $\mathcal{W}$ explicitly written in terms of the embeddings, one may straightforwardly integrate over $\zeta$ to obtain the two-dimensional energy density $w$.  However, our $\mathcal{W}$ in terms of Biot strains is only implicitly related to the embeddings, either through tensor square root operations or simultaneous consideration of rotation fields.  As we only require a low-order approximation of $w$, these difficulties can be avoided by expanding the integrand~\cite{steigmann2007thin,steigmann2008two}.  
Defining 
\begin{align}
	\mathcal{Z} \equiv \sqrt{A}\,(1 - 2\zeta\bar{H}+\zeta^2\bar{K})\,\mathcal{W}\,,\label{zdef}
\end{align}
The two-dimensional energy density $w$ is, at the order we consider, a sum of stretching $w_s$ and bending $w_b$ terms,
\begin{align}
    w/h &= w_s + h^2w_b + O(\Delta^3,\tilde h\Delta^2,h\bar H\Delta^2, \tilde h^2\Delta, h\bar H\tilde h\Delta, h^2\bar K\Delta, \tilde h^3)\,,     \label{eq:wexp} \\
 w_s &= \mathcal{Z}\big|_{\zeta=0}\,,\nonumber\\
 w_b &= \tfrac{1}{24}\mathcal{Z}''\big|_{\zeta=0}\,.\nonumber
\end{align}
We will find later that to this order, the shell terms in the geometric prefactor in \eqref{zdef} will not appear in any of our energies.  From here onwards, we will refrain from writing long lists of higher-order terms when writing expressions.

The bending term contains through-thickness derivatives of $\mathcal{Z}$. 
Recalling the definition \eqref{eq:sigma} of the Biot stress, and using the chain rule and the fact that derivatives of $\bEB$ and $\vmathbb{U}$ are the same, we may write, to the relevant order, 
 \begin{align}
  \mathcal{Z}'|_{\zeta=0} &= \bSB:\vmathbb{U}'|_{\zeta=0} = (\bSB:\bU_\phi'+\bSB:\phi''\mathbf{N}\mathbf{N})|_{\zeta=0}\,, \nonumber \\
  \mathcal{Z}''|_{\zeta=0} &= (\bSB':\vmathbb{U}'+\bSB:\vmathbb{U}'')|_{\zeta=0}
  = (\bSB':\bU_\phi'+\bSB':\phi''\mathbf{N}\mathbf{N}+\bSB:\bU_\phi''+\bSB:\phi'''\mathbf{N}\mathbf{N})|_{\zeta=0}\,.  \label{eq:zdprime}
\end{align}
However, further simplification is possible using boundary conditions on the top and bottom of the shell, and the three-dimensional balance of linear momentum~\cite{steigmann2012extension, steigmann2013koiter}. 
In our specific case, simple informal arguments may be made as follows.
As there are no normal stresses on the top and bottom of the thin shell, the corresponding normal stresses and their derivatives evaluated at the mid-surface will be small. 
 Additionally, the tangential stresses will be of the order of the mid-surface strain, although their $\zeta$-derivatives are still significant.  Therefore, the only term to retain in \eqref{eq:zdprime} is
\begin{align}
 \mathcal{Z}''|_{\zeta=0} &\approx \bSB':\bU_\phi'\big|_{\zeta=0} \label{zprimes2} \\
   &= \left[(2c_1+c_2)\big(\mathrm{Tr}\,\bU_\phi' + \phi''\big)\mathrm{Tr}\,\bU_\phi'
      -c_2\mathrm{Tr}\,(\bU_\phi')^2  \right]\Big|_{\zeta=0}\,,
      \label{eq:zprime3}
\end{align} 
using $(i_1^{\bEB})' = [\mathrm{Tr}(\bU_\phi') + \phi'']|_{\zeta=0}$. Note that $\phi''|_{\zeta=0}=2\alpha_2$.  The requirement that the normal stress and its first derivative are effectively zero will later determine the values of the coefficients $\alpha_1$ and $\alpha_2$.   

To compute the $\zeta$-derivative of the lateral stretch in \eqref{zprimes2}, we recall the assumption presented in Section \ref{sec:def} on the smallness of the $\zeta$-derivative of the rotation, which implies that $(\bg_\alpha\bG^\alpha)'|_{\zeta=0} \approx \bQ\cdot\bU'_\phi |_{\zeta=0}$ and $(\bG^\alpha\bg_\alpha)'|_{\zeta=0} \approx \bU'_\phi\cdot\bQ^\top|_{\zeta=0}$.  Building the deformation gradient requires the basis vectors~\cite{Flugge72}
\begin{align}
  \bG_\alpha = \bA_\alpha+\zeta d_\alpha\mathbf{N} =\,\, &(\bA_\beta\bA^\beta-\zeta\bar{\mathbf{b}})\cdot\bA_\alpha\,, \\
  \bG^\alpha = \quad\quad\quad\quad\quad\quad\,\, &(\bA_\beta\bA^\beta+ \zeta\bar{\mathbf{b}})\cdot\bA^\alpha + O(h^2\bar{\mathbf{b}}^2) \,, \\
  \bg_\alpha =\, \ba_\alpha+\phi d_\alpha\mathbf{n}\,\, =\,\, &(\ba_\beta\ba^\beta-\phi\mathbf{b})\cdot\ba_\alpha\,.
\end{align}
As $\phi$ depends on $\zeta$ through \eqref{eq:phi}, while all the vectors are only defined on the mid-surface, we obtain 
\begin{align}
  (\bg_\alpha\bG^\alpha)'|_{\zeta=0} = -\alpha_1\mathbf{b}\cdot\ba_\alpha\bA^\alpha+\ba_\alpha\bA^\alpha\cdot\bar{\mathbf{b}}\,.
   \label{eq:uphiprime}
\end{align}
We now take care to write the derivative in a manifestly symmetric way, 
\begin{equation}
    \bU_\phi'\big|_{\zeta = 0} = \tfrac{1}{2}\left[\mathbf{Q}^\top\cdot(\bg_\alpha\bG^\alpha)' + (\bG^\alpha\bg_\alpha)'\cdot\mathbf{Q}\right]\big|_{\zeta=0} =
    -\textrm{sym}\left[(\alpha_1\mathbf{Q}^\top\cdot\mathbf{b}\cdot\mathbf{Q}
    -\bar{\mathbf{b}})\cdot\bU\right]\,.\label{symmetricUprime}
\end{equation}
This expression is worthy of comment for two reasons, whose consequences will both propagate forward into our final bending measure. One is that we have avoided deriving an unsymmetric quantity such as Atluri's \cite{atluri1984alternate} 
$\bQ^\top\cdot\mathbf{b}\cdot\bQ\cdot\bU$.
The other is that the referential curvature tensor appears dotted with the stretch, in contrast to its solitary uncoupled appearance in the entirely \emph{ad hoc} postulated forms for curved rods and shells in  \cite{Antman68-2, Reissner72, WhitmanDeSilva74, atluri1984alternate, pietraszkiewicz2008determination, KnocheKierfeld11}. We note also that in independent-rotation shell theories \cite{sansour1992exact, wisniewski1998shell, wisniewski2010finite}, it is useful to consider both symmetric and antisymmetric parts of a ``relaxed'' deformation measure $\bQ^\top\cdot\bg_i\bG^I$ and its derivatives.  When constraining such a theory to conventional elasticity, the symmetric part naturally recovers the right stretch tensor. 

While the appearance of the rotations in \eqref{symmetricUprime} may appear complex and intimidating, we will eventually see that no explicit use of rotation will be required to compute energies, forces, or torques in a shell.  The presence of some sort of two-point tensor is necessary to meaningfully relate the present and referential curvatures $\mathbf{b}$ and $\bar{\mathbf{b}}$.   
Note that our notation is not equivalent to that employed in, for example, \cite{Armon11} and \cite{Pezzulla17}, in which two tensors living in the same tangent space are directly subtracted from each other and only one of them is actually a curvature tensor. 

The traces appearing in \eqref{eq:zprime3} are
\begin{align}
	\mathrm{Tr}\,\bU_\phi'\big|_{\zeta=0} &=
  \bU_\phi':\bA_\alpha\bA^\alpha\big|_{\zeta=0}
  =  -\textrm{Tr}\,\textrm{sym}\left[(\alpha_1\bQ^\top\cdot\mathbf{b}\cdot\bQ-\bar{\mathbf{b}})\cdot\bU\right]
      =  -\textrm{Tr}\,\textrm{sym}\left[\bV\cdot(\alpha_1\mathbf{b}-\bQ\cdot\bar{\mathbf{b}}\cdot\bQ^\top)\right] \nonumber \\
      &\qquad\qquad\qquad\quad\;\;= -\alpha_1\textrm{Tr}\,\textrm{sym}(\bV\cdot\mathbf{b})+\textrm{Tr}\,\textrm{sym}(\bar{\mathbf{b}}\cdot\bU)\,,
  \label{trace} \\
	\mathrm{Tr} (\bU_\phi')^2 \big|_{\zeta=0} &=
	  \bU_\phi':\bU_\phi' \big|_{\zeta=0}
  \quad\;= \textrm{Tr}\,\textrm{sym}^2\left[(\alpha_1\bQ^\top\cdot\mathbf{b}\cdot\bQ-\bar{\mathbf{b}})\cdot\bU\right]  
      \,= \textrm{Tr}\,\textrm{sym}^2\left[\bV\cdot(\alpha_1\mathbf{b}-\bQ\cdot\bar{\mathbf{b}}\cdot\bQ^\top)\right]  
      \,. \label{tracesq}
\end{align}
The symmetrization in \eqref{trace} is redundant but retained for emphasis. 
It remains to determine the coefficients $\alpha_1$ and $\alpha_2$ appearing in \eqref{eq:zprime3} and (\ref{trace}-\ref{tracesq}) for this isotropic quadratic-Biot material. 
In the classical Kirchhoff-Love treatment, they are taken as unity and zero, respectively.

\subsection{Coefficients of $\phi$}
\label{sec:coef}

The first coefficient is determined by the requirement that $\bSB\cdot\mathbf{N} |_{\zeta=0} = \mathbf{0}$.
From \eqref{eq:sigma} and \eqref{eq:biot-exp} we have \mbox{$i_1^{\bEB}  |_{\zeta=0} = \mathrm{Tr}\left(\bU+\alpha_1\mathbf{N}\mathbf{N}-\mathbf{I}\right)$} and
\begin{align}
    \alpha_1 &= 1-\beta\,\textrm{Tr}\,(\bU-\bA_\alpha\bA^\alpha)\,,    \label{eq:a1}\\
    \beta &\equiv 1+c_2/2c_1 \,, \label{eq:beta}
\end{align}
where $\bU - \bA_\alpha\bA^\alpha$ is the surface Biot strain whose invariants are identical to those of the surface Bell strain $\bV - \ba_\alpha\ba^\alpha$. (Note that $\mathrm{Tr}\,\bI = 3$, $\mathrm{Tr}\,\bN\bN = \mathrm{Tr}\,\bn\bn  = 1$, and $\mathrm{Tr}\,\bA_\alpha\bA^\alpha = \mathrm{Tr}\,\ba_\alpha\ba^\alpha = 2$).
This is equivalent to minimizing $w_s$ with respect to $\alpha_1$.  Recall that $c_1$ can only vanish if $c_2$ vanishes, which combination implies no elastic energy whatsoever.

Note that while the full form \eqref{eq:a1} will be used in the stretching energy, when incorporating $\alpha_1$ into the bending energy the $\beta$ term proportional to the surface strain provides a higher-order correction that we neglect.  One way to express this is that the Poisson effect makes an important correction to the stretching content but not to the bending, which is why it was acceptable to ignore it in the physical argument associated with Figure 2 of the companion paper \cite{vitral2021dilation}.

The second coefficient, which appears only in the bending content, is determined by the requirement that \mbox{$\bSB'\cdot\mathbf{N} |_{\zeta=0} = \mathbf{0}$}.  
From the derivative of \eqref{eq:sigma} and \eqref{trace} we have
\begin{align}
	\alpha_2 &= \frac{\beta}{2}\left[ \alpha_1\textrm{Tr}\,\textrm{sym}(\bV\cdot\mathbf{b})-\textrm{Tr}\,\textrm{sym}(\bar{\mathbf{b}}\cdot\bU) \right] \,, \nonumber \\
	&\approx \frac{\beta}{2}\left[ \textrm{Tr}\,\textrm{sym}(\bV\cdot\mathbf{b})-\textrm{Tr}\,\textrm{sym}(\bar{\mathbf{b}}\cdot\bU) \right] \,. \label{eq:alpha2}
\end{align}
Arguments from~\cite{steigmann2010applications} may be used to show that this is equivalent to minimizing the $\bSB':\vmathbb{U}'$ term from $w_b$ with respect to $\alpha_2$ as in~\cite{ozenda2021kirchhoff}.

At this point we observe that the generalized Kirchhoff-Love derivation has, to the order we consider, simply reproduced the coefficients we would have obtained through the classical Kirchhoff-Love derivation with its inconsistent assumptions combining plane stress force balance and plane strain kinematics.

\section{Energies for shells}\label{sec:energy}

\subsection{Stretching energy} 

As expected, the stretching content $w_s$ takes a simple form in terms of the mid-surface Biot strain $\bU - \bA_\alpha\bA^\alpha$ or Bell strain $\bV - \ba_\alpha\ba^\alpha$.  Begin with
\begin{align}
	w_s &= \big[c_1(i_1^{\bEB})^2+c_2i_2^{\bEB}\big]\big|_{\zeta=0} 
	= c_1\Big[ \beta\,\mathrm{Tr}^2\bEB +(1-\beta)\mathrm{Tr}\,\bEB^2 \Big]\Big|_{\zeta=0}  \,, \label{eq:ws} 
\end{align}
and recall from \eqref{eq:beta} that $\beta \equiv 1+c_2/2c_1$. 
Using
\begin{align}
	\mathrm{Tr}\,\bEB \big|_{\zeta=0} &= \mathrm{Tr}\left(\bU+\alpha_1\mathbf{N}\mathbf{N}-\mathbf{I}\right) 
	= (1-\beta)\,\textrm{Tr}\,(\bU-\bA_\alpha\bA^\alpha) \,, \\
	\mathrm{Tr}\,\bEB^2 \big|_{\zeta=0} &= \mathrm{Tr}\left( \left[\bU-\bA_\alpha\bA^\alpha\right]^2 + \left[\left(\alpha_1-1\right)\bN\bN\right]^2 \right)	
	= \mathrm{Tr}\left(\bU-\bA_\alpha\bA^\alpha\right)^2 + \beta^2\,\mathrm{Tr}^2\left(\bU-\bA_\alpha\bA^\alpha\right)
	\,,
\end{align}
we obtain 
\begin{align}
	w_s &= \frac{-c_2}{2}\Big[ \beta\,\mathrm{Tr}^2(\bU-\bA_\alpha\bA^\alpha) +\mathrm{Tr}\,(\bU-\bA_\alpha\bA^\alpha)^2 \Big]  =  \frac{-c_2}{2}\Big[ \beta\,\mathrm{Tr}^2(\bV-\ba_\alpha\ba^\alpha) +\mathrm{Tr}\,(\bV-\ba_\alpha\ba^\alpha)^2 \Big] \,. \label{eq:ws2}
\end{align}
For an incompressible neo-Hookean material with $\mathcal{W}_{nH} \propto \textrm{Tr}\,\vmathbb{C}=\textrm{Tr}\,\vmathbb{U}^2$, only $\mathrm{Tr}\,\bU^2$ would appear in the energy.

\subsection{Bending energies}\label{sec:lurie}

The same tensor measures of bending appear in all of the quantities computed in \eqref{trace}, \eqref{tracesq}, and \eqref{eq:alpha2} that form the building blocks of plate and shell bending energies.
Setting $\alpha_1 = 1$ at our order of approximation of the bending terms, these present and referential measures are
\begin{align}
  \ant_{\mathrm{shell}}=\textrm{sym}[\bV\cdot(\mathbf{b}-\bQ\cdot\bar{\mathbf{b}}\cdot\bQ^\top)]\,,   \label{eq:measure-s} \\ 
  \bar{\ant}_{\mathrm{shell}} = \textrm{sym}[(\bQ^\top\cdot\mathbf{b}\cdot\bQ-\bar{\mathbf{b}})\cdot\bU]\,.  \label{eq:measure-sref}
\end{align}
These share the same invariants.  They are objective, in that under a superimposed rotation $\tilde \bQ$ of the present configuration the present tensor $\ant_{\mathrm{shell}} \rightarrow \tilde\bQ^\top\cdot\ant_{\mathrm{shell}}\cdot\tilde\bQ$ and $\bar{\ant}_{\mathrm{shell}}$ does not change, while under a superimposed rotation $\tilde{\bar\bQ}$ of the reference configuration the present tensor $\ant_{\mathrm{shell}}$ does not change while the referential tensor
 $\bar{\ant}_{\mathrm{shell}} \rightarrow \tilde{\bar\bQ}^\top\cdot\bar{\ant}_{\mathrm{shell}}\cdot\tilde{\bar\bQ}$. 
As mentioned previously, the appearance of the rotation $\bQ$ and its inverse $\bQ^\top$ in (\ref{eq:measure-s}-\ref{eq:measure-sref}) allows both curvature tensors to be properly represented in either a present or referential setting.  The same relations hold between the stretches, namely $\bV = \bQ\cdot\bU\cdot\bQ^\top$ and $\bU = \bQ^\top\cdot\bV\cdot\bQ$, as these are present and referential representations of the strain. 
Note that, for example, the referential representation of the curvature tensor of the present surface $\bQ^\top\cdot\mathbf{b}\cdot\bQ$ is an entirely different tensor than both the curvature tensor of the referential surface $\bar{\mathbf{b}}$ as well as the tensor $b_{\alpha\beta}\bA^\alpha\bA^\beta$ used by ESK \cite{efrati2009elastic}, although it coincides with the latter in the special case of a mid-surface isometry.

The unsymmetric tensor $\bQ^\top\cdot\bb\cdot\bQ\cdot\bU$ was suggested by Atluri~\cite{atluri1984alternate} based on calculations related to the derivation of stress and strain measures for variational formulations of shells.   
 However, both Atluri ~\cite{atluri1984alternate} and Pietraszkiewicz~\cite{pietraszkiewicz2008determination} propose subtracting the referential curvature in an \emph{ad hoc} shell bending measure of the form $\bQ^\top\cdot\bb\cdot\bQ\cdot\bU - \bar \bb$.
The new measures (\ref{eq:measure-s}-\ref{eq:measure-sref}) arise naturally from the derivatives of stretch whose computation is necessary for the reduction of a quadratic-Biot energy, generalizing the one-dimensional and axisymmetric derivations in \cite{IrschikGerstmayr09} and \cite{OshriDiamant17}. 

Employing \eqref{trace}, \eqref{tracesq}, and \eqref{eq:alpha2} to evaluate \eqref{eq:zprime3}, we obtain the bending content in the simple form 
\begin{align}
  w_b &= \frac{-c_2}{24}\left[ \beta\,\textrm{Tr}^2\ant_{\mathrm{shell}} +\textrm{Tr}\,\ant_{\mathrm{shell}}^2 \right] \qquad\quad = \frac{-c_2}{24}\left[ \beta\,\textrm{Tr}^2\bar\ant_{\mathrm{shell}} +\textrm{Tr}\,\bar\ant_{\mathrm{shell}}^2 \right]  \label{eq:wb} \\
          &=  \frac{-c_2}{24}\left[ (1+\beta)\textrm{Tr}^2\ant_{\mathrm{shell}} -2\textrm{Det}\ant_{\mathrm{shell}} \right] = \frac{-c_2}{24}\left[ (1+\beta)\textrm{Tr}^2\bar\ant_{\mathrm{shell}} -2\textrm{Det}\bar\ant_{\mathrm{shell}} \right] \,.
     \label{eq:wb2}
\end{align}
For an incompressible neo-Hookean material, only $\mathrm{Tr}\,\bar\ant_{\mathrm{shell}}^2$ would appear in the energy. 

Any deformation that leaves the bending measure $\ant_{\mathrm{shell}}$ invariant may be defined as a ``pure stretching'' deformation that affects only $w_s$.  By contrast, isometric deformations of the mid-surface are ``pure bending'' that affects only $w_b$.  For such deformations, 
$\bU = \bA_\alpha\bA^\alpha$, $\bV = \ba_\alpha\ba^\alpha$, and $\textrm{Tr}\ant_{\mathrm{shell}} = 2(H-\bar H)$.  However, as will be seen in Section \ref{sec:field}, this definition of pure bending is not the expected response of a curved shell to a pure applied moment. 

Consider the one-dimensional case of the energy \eqref{eq:wb}, corresponding to either unidirectional curvature of a shell, or a naturally-curved rod (beam), with rest mid-line curvature $\mathrm{Tr}\bar\bb = \mathrm{Tr}(\bQ\cdot\bar\bb\cdot\bQ^\top) = \bar \kappa$, deformed so as to have mid-line curvature $\mathrm{Tr}\bb = \kappa$ and mid-line stretch $\mathrm{Tr}\bV = \mathrm{Tr}\bU = \lambda$.
The invariants in the energy reduce to a single quantity 
\begin{align}
	\mathrm{Tr} \ant_{\mathrm{shell}} = \lambda\left(\kappa - \bar\kappa\right)\,.\label{eq:bend1d}
\end{align}
The bending measure $\lambda\kappa$ for a naturally-straight rod has been employed for over half a century, beginning with the work of Antman \cite{Antman68-2, Reissner72, WhitmanDeSilva74}. However, these authors and others \cite{KnocheKierfeld11} propose the one-dimensional or axisymmetric measures $\lambda\kappa - \bar \kappa$, which is the one-dimensional form of the measures proposed by Atluri ~\cite{atluri1984alternate} and Pietraszkiewicz~\cite{pietraszkiewicz2008determination}.  Another common choice is the difference in curvatures $\kappa-\bar{\kappa}$.  
The difference between these quantities is illustrated in their corresponding definitions of ``pure stretching''.  As discussed in the companion paper \cite{vitral2021dilation}, the measure \eqref{eq:bend1d} does not change when a naturally-curved body is extended along itself in its reference configuration, such that $\kappa = \bar\kappa$ and the tangential stretch is uniform across the thickness of the body.

\subsubsection{Plate bending energy}\label{sec:plate}

Certain things simplify when $\bar{\mathbf{b}}=\mathbf{0}$.
The measures $\ant_{\mathrm{shell}}$ and $\bar{\ant}_{\mathrm{shell}}$ reduce to
\begin{align}
	\ant &= \textrm{sym}(\bV\cdot\mathbf{b})\,, \label{eq:measure}\\
	\bar{\ant} &= \textrm{sym}(\bQ^\top\cdot\mathbf{b}\cdot\bQ\cdot\bU)\,. \label{eq:measureref}
\end{align}
The apparent asymmetry in these expressions is a purely notational issue arising from our use of the present curvature $\mathbf{b}$ in both, rather than designating a symbol for the referential version of this tensor $\bQ^\top\cdot\mathbf{b}\cdot\bQ$. 

The plate bending measures (\ref{eq:measure}-\ref{eq:measureref}) are invariant under spatial dilations
 $\bx \rightarrow D\left(\bx - \bx_c\right)$, $D$ and $\bx_c$ constant, which transform $\ba_\alpha \rightarrow D\ba_\alpha$, $\ba^\alpha \rightarrow D^{-1}\ba^\alpha$, and conserve $\bn \rightarrow \bn$, the unit normal perpendicular to these. 
Thus, the rotation $\bQ\rightarrow\bQ$ is conserved while $\bV = \ba_\alpha\bA^\alpha\cdot\bQ^\top \rightarrow D\bV$ and $\bU = \bQ^\top\cdot\ba_\alpha\bA^\alpha \rightarrow D\bU$, and $\mathbf{b} = d_\beta \ba_\alpha\cdot\bn\,\ba^\alpha\ba^\beta \rightarrow  D^{-1}\mathbf{b}$, so that any bilinear product of stretch and curvature is conserved.  Thus, any plate bending energy constructed from invariants of \eqref{eq:measure} or \eqref{eq:measureref} will be dilation-invariant, with dilations superimposed on any deformed surface constituting purely stretching contributions to the elastic energy. 
This generalizes the observations in \cite{WoodHanna19} on the energy of plate-like rings and the one-dimensional measure $\lambda\kappa$ corresponding to the plate case of \eqref{eq:bend1d}.   

Under isometric deformations of the mid-surface, we have $\ant = \bb$ and thus the geometric quantities $\textrm{Tr}\ant = 2H$ $\textrm{Det}\ant = K = 0$ will appear in the energy density obtained from the special case of \eqref{eq:wb2} for isometric plates.  
In general, $\textrm{Det}\ant = \mathrm{Det}\, \bV\, \mathrm{Det}\, \mathbf{b} = JK$ where the Jacobian determinant $J = \sqrt{a/A}$ and $\textrm{d}a = \sqrt{a}\,d\eta^1d\eta^2$ is the present area form.  Thus, part of the plate bending energy is a purely geometric quantity, $\int\! \textrm{d}A\,\mathrm{Det} \ant = \int \textrm{d}A\, JK = \int\! \textrm{d}a\, K$, which is equal to a boundary term plus a topological invariant.

\subsubsection{Comparison with the discrete SN energy}
\label{sec:disc}
  
Numerical implementation of energies such as \eqref{eq:ws2} and \eqref{eq:wb} may be achieved either through finite element or molecular dynamics methods, the latter offering an easier route if a simple algorithm is available.  While we do not offer such an algorithm here, we observe first that the computation of stretches in the stretching energy may be easily obtained from interbead distances, and second that a discrete model commonly used in soft matter applications actually generates part of the plate bending energy.
The latter is a fortunate surprise, as the model was never proposed as a discretization of a dilation-invariant plate energy but rather incorrectly put forward, and is often employed, as a discretization of an energy based on squared curvature. 

Seung and Nelson~\cite{seung1988defects} and many subsequent authors employ an energy of the form
\begin{equation}
    \mathcal{E}_{SNb} \propto \sum_{\langle a,\,b \rangle}(\mathbf{n}_a-\mathbf{n}_b)^2\,,
    \label{eq:bdisc}
\end{equation}
which sum of normals is taken over nearest-neighbor facets in a triangular lattice.  As the bonds of this lattice have unit reference length $\|\bX_a-\bX_b\| =1$, the summand in \eqref{eq:bdisc} represents the square of $(1/\sqrt{3})(\mathbf{n}_a-\mathbf{n}_b)/(\bX_a-\bX_b)$, which is $(1/\sqrt{3})$ times the discretization of the projection of the referential gradient of the normal $\bar\nabla\bn$ on a unit vector in the direction between the triangle centers.  This should be contrasted with $(\mathbf{n}_a-\mathbf{n}_b)/(\bx_a-\bx_b)$, the same operation on the gradient $\nabla\bn$, which is the relevant quantity for generating geometric curvature terms. 
Note the analogous distinction whereby differences in positions $\bx$ (with unit referential distance) are discretizations of components of the referential gradient of displacement, rather than of the present gradient of displacement, which is just the identity tensor. 
The authors~\cite{seung1988defects} 
mistakenly assume that $\mathbf{n}_a-\mathbf{n}_b$ is the discretization of the present gradient of the normal $\nabla\bn$ rather than the referential $\bar\nabla\bn$,  and thus claim that~\eqref{eq:bdisc} is the discretization of a Helfrich energy $\propto \mathcal{E}_{H} = \int_{\mathcal{S}}\!\textrm{d}a\,(2 H^2 - K)$. 
This conclusion is erroneous, both for the misconception about the gradient, as well as the fact that the sum over triangles is not an area-weighted integral $\int \!\textrm{d}a$ but a referential (mass-weighted) integral $\int \!\textrm{d}A = \int \!J^{-1}\textrm{d}a$.    
Other related issues were discussed in \cite{schmidt2012universal}. 
 
Following similar lines as in~\cite{seung1988defects} to interpret the sum over a triangular lattice, the continuum limit of the SN energy is actually
\begin{align}
	\textrm{lim}\, \mathcal{E}_{SNb} \propto \int_{\mathcal{S}}\!\textrm{d}A\,\mathrm{Tr}(\bar{\nabla}\mathbf{n}\cdot\bar{\nabla}\mathbf{n}^\top)
	    = \int_{\mathcal{S}}\!\textrm{d}A\,\textrm{Tr}({\bV^2\cdot\mathbf{b}^2})\,.
    \label{eq:ndisc}
\end{align}
This is one of the dilation-invariant terms appearing in the plate form of the energy derived in this paper. 
The term $\bar{\nabla}\mathbf{n}^\top\cdot\bar{\nabla}\mathbf{n}$, recently proposed by Virga~\cite{virga2021pure} as a measure of pure bending, would provide the same trace. 
A reduction of an incompressible neo-Hookean energy provides a single term of the form $\textrm{Tr}(\mathbf{V}^2\cdot\mathbf{b}^2)+\textrm{Tr}(\bV\cdot\mathbf{b}\cdot\bV\cdot\mathbf{b})$.
Any of these are single-term dilation-invariant plate energies, lacking one of the two parameters available in a general isotropic quadratic energy. 

Finally, we should emphasize that a na{\"{\i}}ve extension of \eqref{eq:bdisc} to include referential normals will not produce a shell energy with the properties of that derived in this paper.


\section{Field equations and boundary conditions}
\label{sec:field}


To derive field equations and boundary conditions, we modify the approach found in Wi{\'s}niewski \cite{wisniewski1998shell} that separates the variation into pieces associated with the variation of position $\delta\bx$, and a pseudovector $\delta\boldsymbol\theta$ that characterizes variation of rotation through the relation
 $ -\bQ\cdot\delta\bQ^\top\cdot() = \delta\bQ\cdot\bQ^\top\cdot() \equiv \delta\boldsymbol\theta\times ()$. 
However, we do not employ his kinematic assumption to perform the variation of curvature, which appears to have led to neglect of the normal component of the resulting force in~\cite{wisniewski1998shell}.  

We evaluate the variation of the energy
\begin{equation}
  \mathcal{E} = \int_{\mathcal{S}}\! \mathrm{d}A\,(hw_s+h^3w_b)\,,
\end{equation}
using the referential forms \eqref{eq:ws2} and \eqref{eq:wb},
\begin{align}
	&h\delta w_s \;\;= h\frac{\partial w_s}{\partial\bU}:\delta\bU \qquad\quad\;\;\; \equiv \mathbf{T}:\delta\bU \quad\quad\,= -hc_2\big[ \beta(\mathrm{Tr}\,\bU-2)\,\bA_\alpha\bA^\alpha+(\bU-\bA_\alpha\bA^\alpha) \big]:\delta\bU \,,  \label{eq:dws1} \\
	&h^3\delta w_b = h^3\frac{\partial w_b}{\partial \bar{\ant}_{shell}}:\delta\bar{\ant}_{shell} \equiv \mathbf{M}:\delta\bar{\ant}_{shell} = -h^3\frac{c_2}{12}\big[\beta\mathrm{Tr}\,(\bar{\ant}_{shell})\,\bA_\alpha\bA^\alpha+\bar{\ant}_{shell}\big]:\delta\bar{\ant}_{shell}\,.  \label{eq:dwb1}
\end{align}    
It can be shown \cite{wisniewski1998shell, IrschikGerstmayr09} that $\mathbf{T} = \int^{h/2}_{-h/2}\textrm{d}\zeta\,\bSB$ and $\mathbf{M} = \int^{h/2}_{-h/2}\textrm{d}\zeta\,\zeta\bSB$. 
Clearly $\mathbf{T}$ is linear in the surface Biot strain $\bU-\bA_\alpha\bA^\alpha$ and $\mathbf{M}$ is linear in the bending measure $\bar{\ant}_{shell}$ but, as briefly explained in the companion paper \cite{vitral2021dilation}, the important property of linearity in the resulting field equations depends on whether the variations $\delta \bU$ and $\delta \bar{\ant}$ themselves introduce any additional nonlinearities.  
Energies based on invariants of geometric curvature or Green-Lagrange strain, for example, will result in nonlinear coupled field equations, while an energy built from Biot strains will lead to forces and moments linear in the stretching and bending measures derived therefrom \cite{IrschikGerstmayr09, OshriDiamant17, WoodHanna19}. 

Using $\mathbf{U} = \mathbf{Q}^\top\cdot\mathbf{a}_\alpha\mathbf{A}^\alpha$ and the fact that $\delta \ba_\alpha = \delta d_\alpha\bx = d_\alpha\delta\bx$, the variation of the stretching content \eqref{eq:dws1} may be written
\begin{align}
  h\delta w_s &= \mathbf{T}:(\delta\bQ^\top\cdot\mathbf{a}_\alpha\mathbf{A}^\alpha+\bQ^\top\cdot\delta\mathbf{a}_\alpha\mathbf{A}^\alpha) \nonumber \\
&= -\bQ\cdot\mathbf{T}\cdot\mathbf{A}^\alpha\cdot\delta\boldsymbol\theta\times\mathbf{a}_\alpha+\bQ\cdot\mathbf{T}\cdot\bA^\alpha:\delta\ba_\alpha \nonumber \\
&= 
  -\mathbf{a}_\alpha\times(\bQ\cdot\mathbf{T}\cdot\mathbf{A}^\alpha)\cdot\delta\boldsymbol\theta
  +\bar{\nabla}_\alpha(\delta\mathbf{x}\cdot\bQ\cdot\mathbf{T}\cdot\bA^\alpha) 
  -\bar{\nabla}_\alpha(\bQ\cdot\mathbf{T}\cdot\bA^\alpha)\cdot\delta\mathbf{x}\,.
  \label{eq:dws2}
\end{align}
Noting that $\bar{\ant} = \textrm{sym}(\bQ^\top\cdot b^\alpha_\beta\,\ba_\alpha\bA^\beta-\bar{\mathbf{b}}\cdot\bU)$, the variation of the bending content \eqref{eq:dwb1} may be written
\begin{align}
  h^3\delta w_b &= \mathbf{M}:\textrm{sym} \big[ \delta\bQ^\top\cdot b^\alpha_\beta\,\ba_\alpha\bA^\beta
  +\bQ^\top\cdot ( \delta b^\alpha_\beta\,\ba_\alpha\bA^\beta
  + b^\alpha_\beta\,\delta\ba_\alpha\bA^\beta )
  -\bar{\mathbf{b}}\cdot\delta\bU \big]\,.
  \label{eq:dwb2}
\end{align}    
Using the Gauss-Weingarten relations $\nabla_\beta\ba_\alpha = b_{\alpha\beta}\bn$ and $b^\alpha_\beta\ba_\alpha = -d_\beta\mathbf{n}$, and defining the two-point tensor
\begin{align}
	\boldsymbol\mu = \bQ\cdot\mathbf{M} \,, \label{eq:mu}
\end{align}
 such that $\mu^{\gamma\beta}\ba_\gamma = \bQ\cdot\mathbf{M}\cdot\bA^\beta$ and $\mu^{\gamma\beta} = \ba^\gamma\cdot\bQ\cdot\mathbf{M}\cdot\bA^\beta$, the term in \eqref{eq:dwb2} involving variation of the rotation becomes
\begin{align}
  &\mathbf{M}:\textrm{sym}(\delta\bQ^\top\cdot b^\alpha_\beta\,\ba_\alpha\bA^\beta) \nonumber \\
  &= (\bQ\cdot\mathbf{M}\cdot\bA^\beta)\cdot(\delta\boldsymbol\theta\times d_\beta\mathbf{n}) \nonumber \\
    &=\big[\ba_\gamma\times\bar{\nabla}_\beta\mu^{\gamma\beta}\mathbf{n}+\bar{\nabla}_\beta(\mathbf{n}\times\mu^{\gamma\beta}\ba_\gamma)\big]\cdot\delta\boldsymbol\theta -\mu^{\gamma\beta}\bar\nabla_\beta\ba_\gamma\cdot (\delta\boldsymbol\theta\times\bn) \,.
\end{align}
Using $\delta a^{\alpha\beta} = -a^{\alpha\gamma}a^{\beta\eta}\delta a_{\gamma\eta}$ and $\delta b_{\alpha\beta} = \nabla_\beta d_\alpha\delta\mathbf{x}\cdot\mathbf{n}$ (this relation is nontrivial and holds even though the variation doesn't pass through the covariant derivative in general~\cite{Hanna19}), along with the Piola identities $\nabla_\alpha\left[J^{-1} ()^\alpha\right] = J^{-1}\bar\nabla_\alpha ()^\alpha$ or $\bar\nabla_\alpha\left[J()^\alpha\right] = J\nabla_\alpha()^\alpha$, the middle terms in \eqref{eq:dwb2} become
 \begin{align}
 	&\int_\mathcal{S}\!\mathrm{d}A\, \mathbf{M}:\textrm{sym}[ \bQ^\top\cdot ( \delta b^\alpha_\beta\,\ba_\alpha\bA^\beta
  + b^\alpha_\beta\,\delta\ba_\alpha\bA^\beta ) ] \nonumber \\
  &= \int_\mathcal{S}\!\mathrm{d}A\, \big( \mu\indices{_\alpha^\beta}\delta b^\alpha_\beta + \mu^{\gamma\beta}b^\alpha_\beta\ba_\gamma\cdot\delta\ba_\alpha \big) \nonumber \\
  &= \int_{\mathcal{S}}\!\mathrm{d}A\,\Big[
  \bar{\nabla}_\alpha\big(
  \bar{\nabla}_\beta\mu^{\alpha\beta}\mathbf{n}\big) \cdot\delta\mathbf{x}
  +\bar{\nabla}_\alpha\big(
  -\bar{\nabla}_\beta\mu^{\alpha\beta}\mathbf{n}\cdot\delta\mathbf{x}
  +\mu^{\beta\alpha}\mathbf{n}\cdot \delta \ba_\beta \big)
+\mu^{\beta\alpha}\bar\nabla_\alpha\ba_\beta\cdot (\delta\boldsymbol\theta\times\bn)
  \Big]\,.
  \label{eq:dwb4}
 \end{align}
The final term in \eqref{eq:dwb2} involving the referential curvature tensor is treated in a similar manner as the stretching terms, becoming
\begin{align}
\mathbf{M}:\textrm{sym}(-\bar{\mathbf{b}}\cdot\delta\bU) = 
 \ba_\alpha\times (\bQ\cdot\textrm{sym}(\mathbf{M}\cdot\bar{\mathbf{b}})\cdot\bA^\alpha)\cdot\delta\boldsymbol\theta
  -\bar{\nabla}_\alpha(\delta\mathbf{x}\cdot\bQ\cdot\textrm{sym}(\mathbf{M}\cdot\bar{\mathbf{b}})\cdot\bA^\alpha) +  \bar{\nabla}_\alpha(\bQ\cdot\textrm{sym}(\mathbf{M}\cdot\bar{\mathbf{b}})\cdot\bA^\alpha)\cdot\delta\mathbf{x}\,.
  \label{eq:dwb6}
\end{align}

Combining everything, we obtain the variation of the energy
\begin{align}
  \delta \mathcal{E}
  = \int_{\mathcal{S}}\!\mathrm{d}A\,\Big(
  \bar{\nabla}_\alpha\big(\mathbf{f}^\alpha\cdot\delta\mathbf{x}+\mu^{\beta\alpha}\mathbf{n}\cdot \delta\ba_\beta\big)
  -\bar{\nabla}_\alpha\mathbf{f}^\alpha\cdot\delta\mathbf{x}
  -\Big[\ba_\alpha\times\mathbf{f}^\alpha+\bar{\nabla}_\alpha\big(\mu^{\beta\alpha}\ba_\beta\times\mathbf{n}\big)\Big]\cdot\delta\boldsymbol\theta
  \Big)\,,
  \label{eq:variation}
\end{align}
where the conserved force is 
\begin{align}
  \mathbf{f}^\alpha = \bQ\cdot[\mathbf{T}-\textrm{sym}(\mathbf{M}\cdot\bar{\mathbf{b}})]\cdot\bA^\alpha
  -\bar{\nabla}_\beta\mu^{\alpha\beta}\mathbf{n}\,,
  \label{eq:force}
\end{align}
with $\mathbf{T}$ and $\mathbf{M}$ defined in (\ref{eq:dws1}-\ref{eq:dws2}) and $\boldsymbol\mu$ in \eqref{eq:mu}.
The field equation conjugate to the variation of position is the balance of linear momentum
\begin{align}
	\bar\nabla_\alpha\mathbf{f}^\alpha = \mathbf{0} \,,\label{linearmomentumbalance}
\end{align}
which is accompanied by conditions on a boundary $\partial \mathcal{S}$ with tangent normal $\bar\nu_\alpha$, 
\begin{align}
	\bar\nu_\alpha \mathbf{f}^\alpha = \mathbf{0} \, , \\
	\bar\nu_\alpha \mu^{\beta\alpha} \bn  = \mathbf{0} \,.
\end{align}
Note that the boundary condition conjugate to the derivative of the variation of position is not strictly for the moment, although the latter could be derived from it.
The piece of the variation \eqref{eq:variation} conjugate to $\delta\boldsymbol\theta$ directly provides the balance of angular momentum in the traditional form for shells, 
\begin{align}
	\bar{\nabla}_\alpha(\mu^{\beta\alpha}\ba_\beta\times\mathbf{n}) + \ba_\alpha\times\mathbf{f}^\alpha = \mathbf{0} \, , \label{angularmomentumbalance}
\end{align}
while the conserved torque can be identified either by using the balances \eqref{linearmomentumbalance} and \eqref{angularmomentumbalance},  
or by inserting a small rotation $\delta\bx = \epsilon\hat{\bm{\omega}}\times\bx$, $\epsilon\hat{\bm{\omega}}$ constant, into the total boundary term in \eqref{eq:variation}.  Either way one obtains
\begin{align}
  \mathbf{m}^\alpha = \mu^{\beta\alpha}\ba_\beta\times\mathbf{n} + \bx \times\mathbf{f}^\alpha
  \,,
  \label{eq:torque}
\end{align}
such that 
\begin{align}
 \bar{\nabla}_\alpha\mathbf{m}^\alpha = \mathbf{0}\,.
\end{align}
The quantity $\mu^{\beta\alpha} \ba_\beta$ is linear in 
 the bending measure $\bar{\ant}_\mathrm{shell}$, just as the tangential force is linear in 
 the surface Biot strain. 

Consider applying a pure moment to the surface.  For a plate, $\bar\bb=\mathbf{0}$ and it is clear from the form of the tangential part of $\mathbf{f}^\alpha = \mathbf{0}$ that the surface Biot strain will vanish, so that the response will be a mid-plane isometry. 
For a shell, however, the term containing $\bar{\mathbf{b}}$ indicates that the mid-surface will be strained.  
We find that this strain recovers the uniquely defined neutral surface of a naturally-curved beam from classical linear elasticity.
This surface is a property of the structure that does not depend on the magnitude of the applied moment.  For a plate, by symmetry, this must be the mid-surface.  
This result contrasts with those arising from models based on squared curvature or the bending measures derived from Green-Lagrange strains, for which a symmetric plate's midsurface is not neutral, and thus no unique neutral surface exists independently of the applied moment.
For shells, models in which a referential curvature is subtracted from a plate bending measure similar to ours also fail to reproduce the classical neutral surface.

For a thin, curved beam with rectangular cross section, the neutral line is displaced towards the concave side of the beam by a distance $\zeta_\mathrm{n} = h(\frac{h}{12}\bar{\kappa} + O(h^3\bar\kappa^3))$ from the centerline \cite{UguralFenster}.  
Consider the tangential stretch $\lambda(\zeta)$ parallel to the mid-surface; on the neutral line $\lambda(\zeta_n)=1$.  
 In one dimension, a pure moment yields $\mathbf{T} = \mathbf{M}\cdot\bar{\mathbf{b}}$ and thus $\lambda(0) - 1 = \frac{h^2}{12}\lambda(0)(\kappa-\bar{\kappa})\bar{\kappa}$.
 Similarly to \eqref{eq:biot-exp}, we may expand $\lambda(\zeta) \approx \lambda(0)(1-\zeta(\kappa-\bar{\kappa}))$ and set this equal to unity to find the neutral line at $\zeta_\mathrm{n} \approx \frac{h^2}{12}\bar{\kappa}$. 
 
 Given that this neutral surface is a property of the structure, it should be possible to reformulate everything in terms of its embedding rather than that of the mid-surface.


\section{Expressions in terms of metrics and curvatures}
\label{sec:determination}

While formulations of three-dimensional elasticity in terms of stretches and rotations can be difficult or require approximations \cite{vitral2021quadratic}, certain simplifications are possible for two-dimensional bodies that allow us to express the energy purely in terms of the components of the present and referential metric and curvature tensors, which are easily computed from derivatives of position.

The following relations can be obtained either through $\bU\cdot\bU = \bC$ and $\bV\cdot\bV = \bB$ or the Cayley-Hamilton theorem in two dimensions~\cite{hoger1984determination,pietraszkiewicz2008determination}:
\begin{align}
  \bU = \frac{1}{\textrm{Tr}\,\bU}\bigg[\bU\cdot\bU+(\textrm{Det}\,\bU)\,\bA_\alpha\bA^\alpha\bigg] \,,\quad 
  \bV = \frac{1}{\textrm{Tr}\,\bV}\bigg[\bV\cdot\bV+(\textrm{Det}\,\bV)\,\ba_\alpha\ba^\alpha\bigg]\,,
\end{align}
where
\begin{align}
	&\textrm{Det}\,\bU = \textrm{Det}\,\bV =  J = \sqrt{a/A} \,, \label{determinants} \\
	&\textrm{Tr}\,\bU = \textrm{Tr}\,\bV = \sqrt{a_{\gamma\eta}A^{\eta\gamma}+2\sqrt{a/A}}\,,\label{traces}
\end{align}
and thus
\begin{equation}
  U_{\alpha\beta} = \frac{a_{\alpha\beta}+\sqrt{a/A}\,A_{\alpha\beta}}{\sqrt{a_{\gamma\eta}A^{\eta\gamma}+2\sqrt{a/A}}}\,,\quad
  V^{\alpha\beta} = \frac{A^{\alpha\beta}+\sqrt{a/A}\,a^{\alpha\beta}}{\sqrt{a_{\gamma\eta}A^{\eta\gamma}+2\sqrt{a/A}}}\,,
\end{equation}
making use of curious identities between components in different bases, such as $U_{\alpha\beta} U^\beta_\gamma = a_{\alpha\gamma}$ and $V^{\alpha\beta}V_\beta^\gamma = A^{\alpha\gamma}$. 
Recall that $U^\alpha_\beta = A^{\alpha\gamma}U_{\gamma\beta}$ and $V^\alpha_\beta = a_{\beta\gamma}V^{\gamma\alpha}$, so care should be taken when operating on indices in mixed expressions. 

The terms in the stretching content \eqref{eq:ws2} are
\begin{align}
	&\mathrm{Tr}^2(\bU-\bA_\alpha\bA^\alpha) = (U_\alpha^\alpha - 2)^2 \qquad\qquad\quad\;\;\;= (V_\alpha^\alpha - 2)^2 = \mathrm{Tr}^2(\bV-\ba_\alpha\ba^\alpha) \,, \\
	&\mathrm{Tr}(\bU-\bA_\alpha\bA^\alpha)^2 = U^\alpha_\beta U^\beta_\alpha - 2U^\alpha_\alpha + 2 = V^\alpha_\beta V^\beta_\alpha - 2V^\alpha_\alpha + 2 = \mathrm{Tr}(\bV-\ba_\alpha\ba^\alpha)^2\,,
\end{align}
with $U_\alpha^\alpha = V_\alpha^\alpha = \sqrt{a_{\gamma\eta}A^{\eta\gamma}+2\sqrt{a/A}}$
and $U^\alpha_\beta U^\beta_\alpha = V^\alpha_\beta V^\beta_\alpha = a_{\alpha\beta}A^{\alpha\beta}$. 

To obtain the terms in the bending content \eqref{eq:wb} we first recognize that the rotations are acting on surface tensors, so that we need only invert their surface projections.  Using $\bQ^{-1} = \bQ^{\top}$ and~\eqref{eq:polar}, we may write $\bQ^\top\cdot\mathbf{b}\cdot\bQ = U_{\alpha\gamma}b^{\gamma\eta}U_{\eta\beta}\,\bA^{\alpha}\bA^{\beta}$ and $\bQ\cdot\bar{\mathbf{b}}\cdot\bQ^\top = V^{\alpha\gamma}\bar{b}_{\gamma\eta}V^{\eta\beta}\,\ba_{\alpha}\ba_{\beta}$, 
being careful to remember that the reference metric operates on the indices of the $\mathbf{U}$ and $\bar\bb$ components while the present metric operates on $\mathbf{V}$ and $\bb$ (another notational option is to use capitalization on referential indices, as in \cite{vitral2021quadratic}). 
Then the bending measures (\ref{eq:measure-s}-\ref{eq:measure-sref}) are
\begin{align}
	\ant_{\mathrm{shell}} &= \tfrac{1}{2} L\indices{^\alpha_\beta}(\ba_\alpha\ba^\beta + \ba^\beta\ba_\alpha)\,,\\
	L\indices{^\alpha_\beta} &= V^{\alpha\gamma}(b_{\gamma\beta} - V_\gamma^\eta\bar b_{\eta\mu} V^\mu_\beta) = V^{\alpha\gamma} b_{\gamma\beta} - \bar b^\alpha_\gamma V^\gamma_\beta \,,\\
	\bar\ant_{\mathrm{shell}} &= \tfrac{1}{2} \bar L\indices{^\alpha_\beta}(\bA_\alpha\bA^\beta + \bA^\beta\bA_\alpha)\,,\\
	\bar L\indices{^\alpha_\beta} &= (U^\alpha_\mu b^{\mu\eta}U_\eta^\gamma - \bar b^{\alpha\gamma})U_{\gamma\beta} =  U^\alpha_\gamma b^\gamma_\beta - \bar b^{\alpha\gamma} U_{\gamma\beta} \,,
\end{align} 
so that the quantities in the bending content \eqref{eq:wb} are $\mathrm{Tr}^2\ant_{\mathrm{shell}} = L\indices{^\alpha_\alpha}L\indices{^\beta_\beta}$ and 
$\mathrm{Tr}\ant_{\mathrm{shell}}^2  
= \tfrac{1}{2}L\indices{^\alpha^\beta}\big(L\indices{_\alpha_\beta}+L\indices{_\beta_\alpha}\big)$ or the equivalent referential forms, with the same caveats on index manipulation.

Quantities appearing in the field equations include
\begin{align}
  \mu^{\alpha\beta} &= a^{\alpha\gamma}U_{\gamma\eta}M^{\eta\beta}\,,  \\
  \mathbf{f}^{\alpha} &= \mathbf{a}^\beta U_{\beta\gamma}[T^{\gamma\alpha}-\tfrac{1}{2}(M^{\gamma\eta}\bar{b}_\eta^\alpha+\bar{b}^\gamma_\eta M^{\eta\alpha})] + \bar{\nabla}_\beta\mu^{\alpha\beta}\mathbf{n}\,,
\end{align}
where the components of the referential tensors $\mathbf{T}$ and $\mathbf{M}$ can be inferred from (\ref{eq:dws1}-\ref{eq:dwb1}) and other expressions in the current section.
During manipulations, it should be kept in mind that $\mu^{\alpha\beta}$ are components of a two-point tensor of the same character as the rotation or deformation gradient, with a left present index and a right referential index.


\section{Conclusions}\label{sec:conc} 

We have derived stretching and bending energies for isotropic elastic plates and shells by dimensional reduction of a bulk elastic energy quadratic in Biot strains.  
A generalized Kirchhoff-Love kinematics, which reproduces the classical results to low order in stretching and bending, naturally generates new symmetric measures of bending bilinear in stretches and geometric curvatures, through which the bending energy adopts a simple form.  This extends primitive measures for straight rods to plates and naturally-curved rods and shells, and reveals problems with assumed forms for shells in prior literature.  
The bending energy shares properties with both neo-Hookean and discrete SN energies but is fully general at quadratic order in stretch.  The plate form is dilation-invariant.  
Field equations and boundary conditions demonstrate the linearity of force and moment in the resulting measures, and the consequence that the response to a pure moment is a neutral-surface isometry in agreement with classical linear results.
For these two-dimensional elastic bodies, it is possible to represent all the relevant quantities in terms of derivatives of position, although such expressions are not simple.


\section*{Acknowledgments}

This work was supported by U.S. National Science Foundation grant CMMI-2001262.  
We thank E. G. Virga for extensive detailed discussions and for sharing notes on related work.  
We also acknowledge helpful discussions with S. Cheng, P. Plucinsky, and E. Vouga.

\appendix

\bibliographystyle{unsrt}

\end{document}